\documentclass[5p,authoryear]{elsarticle}
\usepackage{amsmath}
\usepackage{amsthm}
\usepackage{aas_macros}
\usepackage{algorithm}
\usepackage{algpseudocode}
\usepackage{hyperref}
\newtheorem{definition}{Definition}
\newtheorem{theorem}{Theorem}

\usepackage{color}
\journal{Astronomy and Computing}

\begin{document}

\begin{frontmatter}

\title{A fast algorithm for identifying Friends-of-Friends halos}

\author[bccp,bids,corauth]{Yu Feng}
\ead{yfeng1@berkeley.edu}

\author[bccp]{Chirag Modi}
\ead{modichirag@berkeley.edu}
\cortext[corauth]{Corresponding author}
\address[bccp]{Berkeley Center for Cosmological Physics Campbell Hall 341, University of California, Berkeley CA 94720}
\address[bids]{Berkeley Institute for Data Science, Doe Library 140, University of California, Berkeley CA 94720}

\begin{abstract}
We describe a simple and fast algorithm for identifying friends-of-friends features and prove its correctness. The algorithm avoids unnecessary expensive neighbor queries, uses minimal memory overhead, and rejects slowdown in high over-density regions. We define our algorithm formally based on pair enumeration, a problem that has been heavily studied in fast 2-point correlation codes and our reference implementation employs a dual KD-tree correlation function code. We construct features in a hierarchical tree structure, and use a splay operation to reduce the average cost of identifying the root of a feature from $O[\log L]$ to $O[1]$ ($L$ is the size of a feature) without additional memory costs. This reduces the overall time complexity of merging trees from $O[L\log L]$ to $O[L]$, reducing the number of operations per splay by orders of magnitude. We next introduce a pruning operation that skips merge operations between two fully self-connected KD-tree nodes. This improves the robustness of the algorithm, reducing the number of merge operations in high density peaks from $O[\delta^2]$ to $O[\delta]$. We show that for cosmological data set the algorithm eliminates more than half of merge operations for typically used linking lengths $b \sim 0.2$ (relative to mean separation). Furthermore, our algorithm is extremely simple and easy to implement on top of an existing pair enumeration code, reusing the optimization effort that has been invested in fast correlation function codes.
\end{abstract}


\end{frontmatter}

\section{Introduction}

Friends-of-Friends clustering (FOF) is a common problem in cosmology for identifying features (clusters, usually called halos or groups) in density fields. Three common uses are 1) to find halos from N-body computer simulations in the 3-dimensional configuration space \citep{1985ApJ...292..371D}; 2) to find sub structures inside halos from N-body computer simulations in the 6-dimensional phase space \citep{2010MNRAS.408.1818W,2013ApJ...762..109B}; 3) and galaxy clusters from observational catalogs \citep{2012MNRAS.420.1861M} in the red-shifted configuration space. To assemble a physical catalog based on the feature catalog from the FOF algorithm, it is typical to prune the features (with some dynamical infall model), and to compute and associate additional physical attributes (e.g. spherical over-density parameters).

FOF algorithms identify features (or clusters) of points that are (spatially) separated by a distance that is less than a threshold (linking length $b$, typically given in units of mean separation between points) and assigns them a common label. A typical algorithm that solves this involves a breadth-first-search (henceforth BFS). During each visit of BFS, a neighbor query returns all of the particles within the linking length of a given particle. The feature label of these neighbors are examined and updated, and the neighbors whose labels are modified are appended to the search queue for a revisit. The first description of the friends-of-friends algorithm with breadth-first-search in the context of astrophysics following this paradigm is by \citep{1983ApJS...52...61G}. A popular implementation is by \cite{KDFOF}, and more recently by \cite{2016MNRAS.459.2118K}.  A naive BFS algorithm queries perform neighbor queries on a point for multiple times, which is an target for optimization. For example, \cite{Kwon:2010:SCA:1876037.1876051} reduce the number of queries by skipping visited branches of the tree.

Another widely used algorithm creates the friends-of-friends features by hierarchical merging \citep[e.g.][]{2005MNRAS.364.1105S}. This was originally used for parallelization on large distributed computer architectures, as it allows a very large concurrency with a simple decomposition of the problem onto spatially disjoint domains. The algorithm is implemented in the popular simulation software GADGET\footnote{though not available in the public version}, but probably existed long before. It has been adopted in many codes, including a publicly available version in the AMR code ENZO \citep{2014ApJS..211...19B}. To improve upon spatial queries, GADGET incrementally increase the linking length with multiple iterations. During each iteration, the algorithm performs a neighbor query on a selected set of points, and merges the proto-features(proto-clusters) hosting these points by updating the labels of all constituent points of these two proto-features. The iterations are repeated till no additional merging is possible, and as a result, multiple neighbor queries on a data point are performed.

In the GADGET implementation, the proto-features are maintained as a forest of threaded trees, where the leaves (points) of any tree are connected by a linked list (hence the name threaded). During a merge operation, two link lists are joined, by traversing to the tail of the shorter linked list and connecting it to the head of the longer linked list. Two additional storage spaces of $O[N]$ are required to keep track of the size of proto-features and the threading linked list. The traverse increases the cost to merge a feature of length $L$ to $O[L \log L]$, which can be a factor of a few more than optimal in terms of wall clock time. This short-coming of a linked list representation is discussed in detail in Section 21 of \cite{Cormen:2009:IAT:1614191}. 

Due to these multiple iterations of the data, each making many expensive spatial queries (that slows down significantly as over-density grows) required in the existing algorithms, FOF has been generally considered a slow algorithm. As a result, algorithms that leads to an exact solution are rarely discussed in any detail in the literature of cosmology and astrophysics, while numerous approximated FOFs have been proposed as better alternatives to trade the speed with accuracy, some with more desirable physical characters (e.g. avoid bridging -- counting nearby halos as one). The general idea of these approximated methods is that accurately tracking the outskirts of halos (features) is not important as it is already dominated by shot-noise in the numerical scheme of solvers. A few examples are improving the speed by using density information \citep{1998ApJ...498..137E}, stochastic sub-sampling \citep{2008ApJ...681.1046L}, and a relaxed linking length \citep{AFOF}. 

Conceptually the FOF problem of cosmology is the same as a well known problem of computer science -- that of identifying the maximum connected components (MCC) from a graph, where the graph is induced from the data set with an adjacent matrix
\[
A(i, j) = 
\begin{cases}
0,& Dist(i, j) > b \\
1,& Dist(i, j) \le b,
\end{cases}
\]
where $b$ is the linking length. Put differently, if there is a path between two points, then they belong to the same feature, which is represented by a disjoint set. This problem is well studied and has a wide range of applications beyond the field of astrophysics. Numerous example implementations are freely available and integrated into machine learning packages. \citep[e.g.][]{Shun:2013:LLG:2517327.2442530}

In this paper we apply well known data structures and algorithms from computer science to derive a fast exact friends-of-friends algorithm that avoids expensive neighbor queries, uses minimal memory overhead, and rejects over-density slow down. 

Our main inspiration is from the dual-tree algorithm introduced by \cite{2001misk.conf...71M}. The dual-tree algorithm efficiently calculates correlation functions by walking two spatial index trees simultaneously and avoids expensive and unnecessary neighbor queries. We use KD-Tree in the example implementation, though this can be replaced with a ball-tree for higher dimensional data and a chaining mesh for low dimensional data to achieve better performance \citep[for the latter, see][]{manodeep_sinha_2016_55161}. Most importantly, the dual-tree algorithm calculates the correlation function with a single pass, enumerating each pair of neighboring points exactly once. Rewriting the FOF algorithm with pair enumeration avoids the repeated neighbor queries in breadth-first-search (BFS) algorithms and the GADGET hierarchical algorithm. 

The main issue in the hierarchical merging algorithm, as pointed above, is the costly hierarchical merging of proto-features. We address this by representing the proto-features with a tree/forest data structure, and apply a splay operation in the merge procedure, which moves recently accessed nodes closer to the root, accelerating root finding operations in the average case \citep[][Section 21]{Cormen:2009:IAT:1614191}. The splay operation was original introduced  by \cite{Sleator:1985:SBS:3828.3835} to balance binary tree structures. In our case, splay reduces the average case complexity to construct a final features of length $L$ to $O[L]$ (as compared to $O[L\log L]$ with a linked list, as implemented in GADGET). It also eliminates the need to use additional $O[N]$ storage space for threading and balancing, resulting an extremely simple implementation. For completeness, we give an intuitive proof of finding correct solution with a single pass of pair enumeration with the splay tree data structure.

To further speed up our algorithm, especially in case of heavily over dense region where spatial queries become increasingly expensive (scaling as $O[((1+\delta) b^3)^2]$ where $b$ is the linking length and $\delta$ is the over density), we implement another important optimization. We show that if two KD-Tree nodes (proto-features) are known to be fully-connected, the nodes need not be further opened and their respective hosting proto-features can be directly merged. This optimization eliminates most of merge operations in dense region and is particularly relevant in high resolution simulations that resolves Kpc scale structures and over-density peaks of $\delta \gg 10^3$ \citep[if we push high resolution simulations such as][to a cosmological volume]{2014MNRAS.445..581H}, though even for current generation of simulations it already reduces the number of merge operations by 20\% to 50\%.

The algorithm can be directly applied as the local section of a parallel friend of friend halo finding routine. Our implementation of the algorithm is available at \url{https://github.com/rainwoodman/kdcount/blob/master/kdcount/kd_fof.c}. We note that our reference dual tree pair enumeration code is not particularly optimized for performance, and hence we rather focus on the theoretical aspects of the algorithm and optimizations in this work. One can easily re-implement our algorithms with existing highly optimized fast correlation function codes to further improve the performance of FOF halo identification on actually problems.

The paper is organized as the following: in Section 2, we define the plain dual-tree friends-for-friends algorithm and prove its correctness; in Section 3, we will discuss the optimization; in Section 4, we perform scaling tests of the algorithm on two realistic cosmological simulation data sets.

\section{Dual tree Friends-of-Friends algorithm}
In this section, we describe our main algorithm, which is based on walking simultaneously two KD-trees that spatially indexes the data set being analyzed.

\begin{definition}
We define a KD-Tree with $M$ nodes as a tuple of ($L[0:M]$, $R[0:M]$, $P[0:M]$), where $L[m]$ is the left child of $m$, $R[m]$ is the right child of $m$, and $P[m]$ is the list of points contained by $m$. We follow the convention that $0$-th node is the root node.
Several operations are also defined:
\begin{itemize}
\item $Dist(i, j) \equiv$ distance between $i$-th and $j$-th point in the dataset. Every time a pair of points are enumerated a $Dist(i, j)$ operation is performed.
\item $MinDist(m, n)$/$MaxDist(m, n)$, the minimal / maximal distance of pairs between $m$-th and $n$-th node;
\item $MinDistB(m, n)$/$MaxDistB(m,n)$, the bounds of minimal / maximal distance between $m$-th and $n$-th node. 
\end{itemize}
\end{definition}
The bounds are quickly computed from the bounding geometry of the KDTree nodes, as in a pair counting algorithm. We use the bound properties 
\[
MinDistB(m, n) < MinDist(m, n)
\]
and
\[
MaxDistB(m, n) > MaxDist(m, n)
\]
to avoid computing the expensive exact minimal / maximal distances. This process is commonly known as pruning.

\subsection{Pair enumeration with KD-Tree}
The FOF algorithm is based on pair enumeration and therefore we first briefly review the relevant pieces of the pair enumeration algorithm for clarity and completeness.

We follow the auto-correlation scenario described in \cite{2001misk.conf...71M} and set both starting location to the root node of the KD-Tree. Here, we simply give an outline of the algorithm and refer the readers to the original reference \citep{2001misk.conf...71M} for proofs and detailed discussions. 

The pseudo code in Algorithm \ref{alg:enum} describes the operation $enum(m, n, b)$ that returns a list of all pairs of data points that are maximally separated by distance $b$ and contained in $m$-th and $n$-th nodes.

\begin{algorithm}
\caption{enum: enumerate the edges that connect a pair of particles with distance less than $b$ with the dual tree method on tree $(L, R, P)$. We have written it as a generative function that yields the pairs to emphasize that the full list does not need to be saved in memory.
}
\label{alg:enum}
\begin{algorithmic}
\Procedure{enum}{$L$, $R$, $P$, $m$, $n$, $b$}
\If{$m$ is a leaf}
   \State{\textbf{Swap} $m$ and $n$}
\EndIf
\If{$\Call{MaxDistB}{m, n} < b$ or $m$ is a leaf}
	\ForAll{$i \in P[m]$, $j \in P[n]$}
    	\If{$\Call{Dist}{i, j} < b$} 
	    	\State{\textbf{yield} i, j}
        \EndIf
    \EndFor
\Else
    \If{$\Call{MinDistB}{m, n} < b$}
        \Comment{enumerate nodes only if they are sufficiently nearby}
		\State{\textbf{yield all} $\Call{enum}{L, R, P, L[m], n, b}$}
        \State{\textbf{yield all} $\Call{enum}{L, R, P, R[m], n, b}$}
    \Else
    	
    \EndIf
\EndIf
\EndProcedure
\end{algorithmic}
\end{algorithm}

\subsection{Friends-of-friends with pair enumeration}

Next we describe the dual tree friends-of-friends algorithm. In addition to the data structures spatially indexing the data points, this algorithm requires us to maintain (both as an output and as a scratch space) an associated array storing the the (proto-)feature labels of the points.
\begin{definition}
We define an array $H[0:N]$ to represent friends-of-friends features, where $N$ is the number of points in the data set. If $p$ is the parent of $i$, we set $H[i] = p$. The root of a (proto-)features satisfies $H[r] = r$ and thus acts as the label of the (proto-)features. 
\end{definition}
$H$ defines a forest, from which we can find the root $r_i = \mathbf{root}(i)$ of the proto-feature hosting $i$ by back-tracing in $H$. The root acts as the label of the proto-features: merging two proto-features with root $r_i, r_j$ is done by setting $H[r_i] \Leftarrow r_j$.

Given the definition of $H$ and the pair enumeration algorithm, we are ready to describe our friends-of-friends algorithm. The algorithm consists of three steps:
\begin{enumerate}
\item At the beginning of the algorithm, we initialize $H[i] = i$, such that each point forms a proto-feature of size $1$ containing the point itself. 
\item We then visit (and merge the proto-features) each pair $(i, j)$ yielded by the pair enumeration algorithm (and hence is closer than the linking length $b$ $\Rightarrow$ belong to the same proto-feature), we find the roots for $i$ and $j$, $r_i$ and $r_j$ by backtracking in $H$ and merge the two proto-features. 
\item At the end of the algorithm, set $H[i] = r_i$.
\end{enumerate}

\begin{figure*}
\includegraphics[scale=0.70]{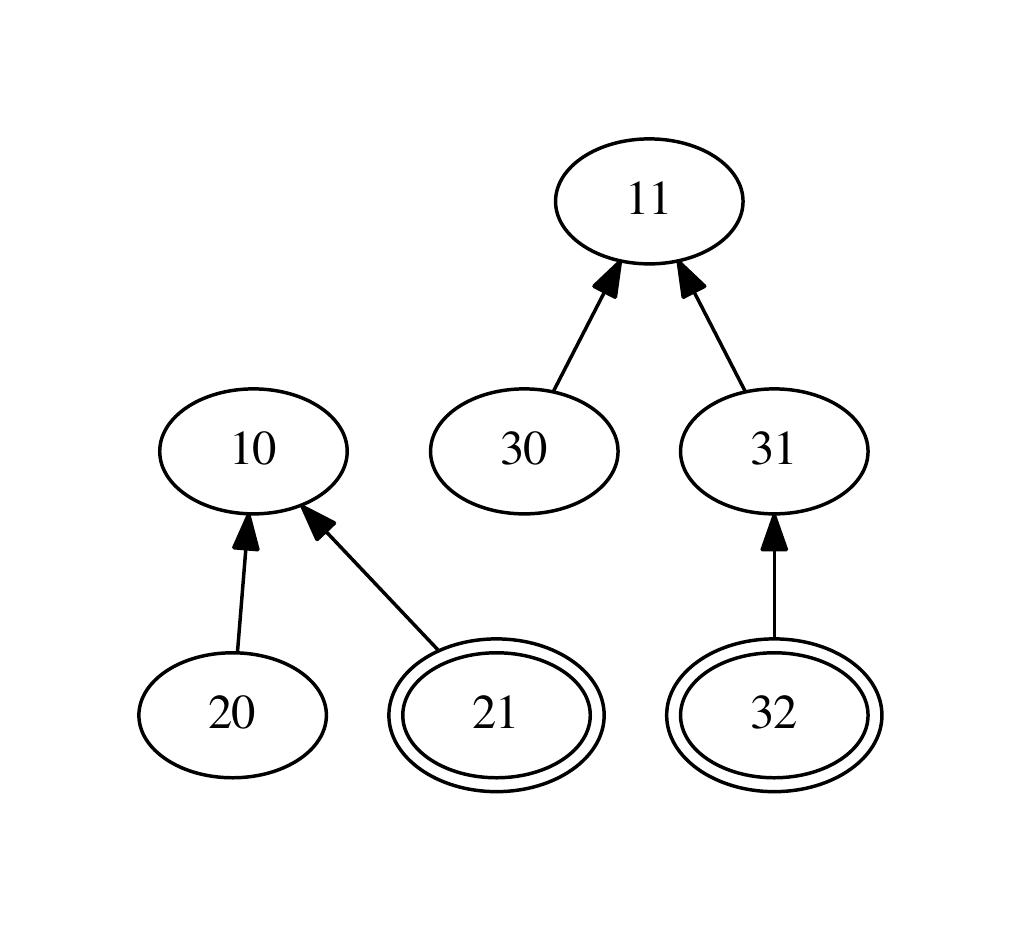}
\includegraphics[scale=0.70]{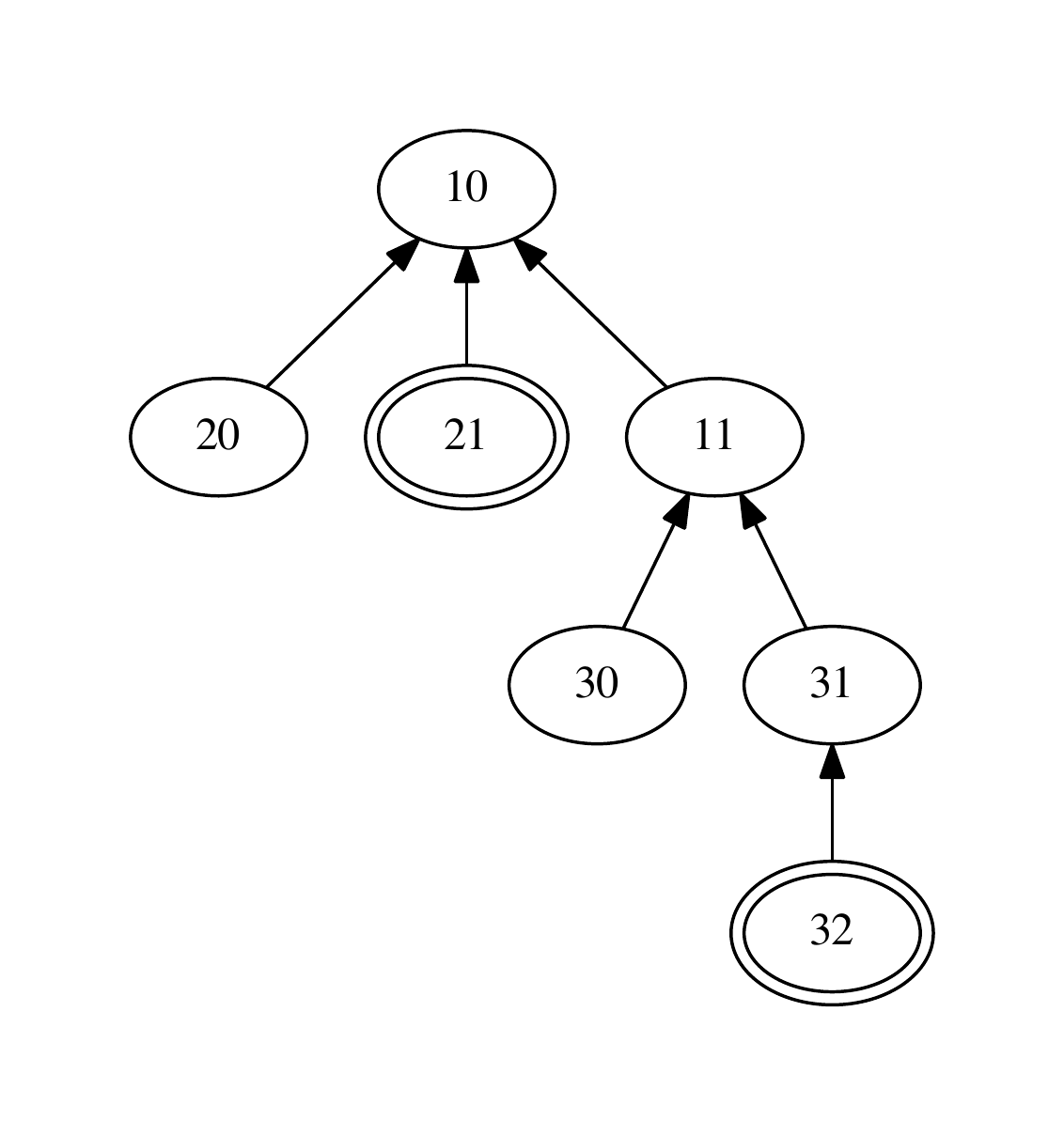}
\caption{Merging two proto-features with the naive method described in Section 2.2. The left panel shows the two proto-features to be merged. The pair of points being visited are marked with double circles. The right panel shows the result after the merge. The depth of the result increases, increasing the cost of finding the root.}
\label{fig:naive-merge}
\end{figure*}

The step is illustrated with an example in Figure \ref{fig:naive-merge}. We see that this naive way of merging proto-features leads to very deep sub-trees with increasingly expensive root-query by back-tracing in $H$. For instance, to merge a proto-feature of length $l$, the root-query takes $O(l)$ time and thus doing this for all the $L$ particles to generate a length $L$ feature can take upto $O(L^2)$ time if the hierarchical tree structure is very unbalanced.

Linked list implementation of FOF algorithms (as done in GADGET) overcomes the depth problem by keeping track of size of proto-feature and during merge operation, it traverses the shorter tree and merges all its leaves directly to the root of longer tree ($H[k] \Leftarrow r_j$ such that $ \mathrm{root}(k) = \mathrm{root}(i)$). This makes the root-query of $O(1)$. However, since the shorter array needs to be traversed to update labels, the overall time complexity to merge a size L tree reduces only to $O(L \log L)$. This can be seen by considering that at every stage, the size of the shorter tree (whose labels $H[i]$ are changed) is at least doubled. Thus, for any point $i$ in the feature of size $L$, the maximum number of times its labels change is $\log L$ and thus for all L points, time taken is $O(L \log L)$. \citep[Also discussed in][Section 21, Data structures for disjoint sets]{Cormen:2009:IAT:1614191}

\subsection{Optimization with splay}
\begin{figure*}
\includegraphics[scale=0.70]{two-proto-features}
\includegraphics[scale=0.70]{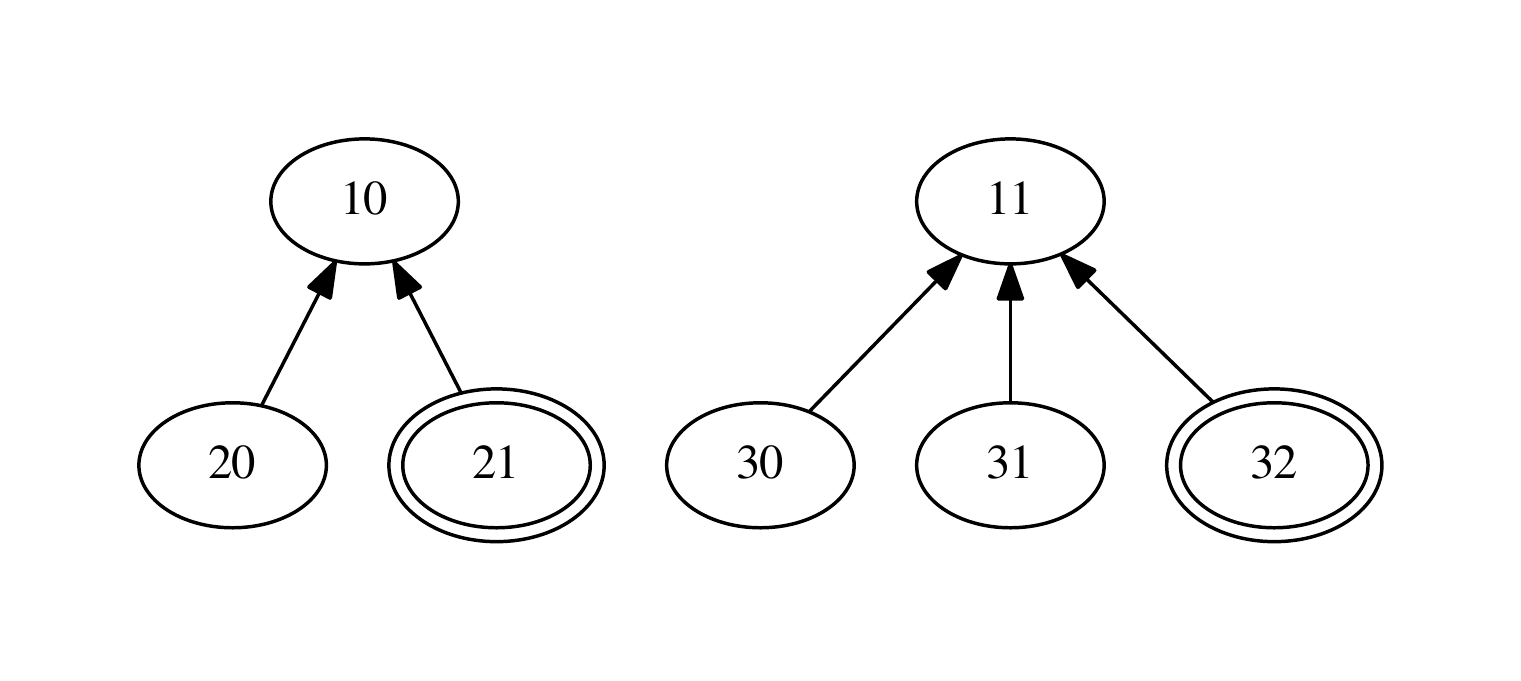}
\caption{Splay operation. The left panel shows the two proto-features to be merged. The right panel shows the result after splay is applied to the data point being visited. Splay flattens the branch being visited on the second tree (root 11), reducing the depth of the tree.}
\label{fig:splay}
\end{figure*}

\begin{figure}
\includegraphics[scale=0.70]{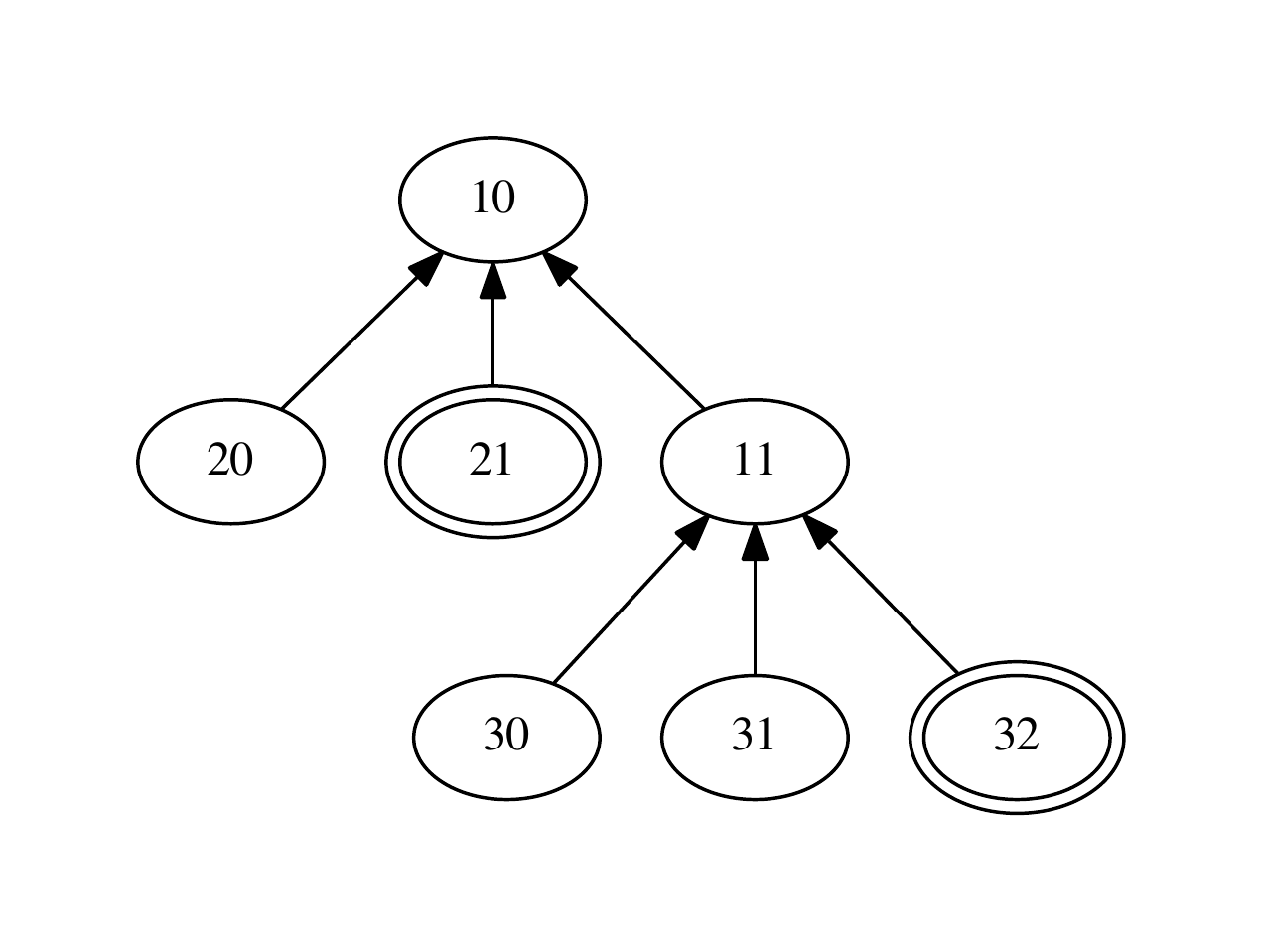}
\caption{Merging the splayed proto-features. Continuing from Figure \ref{fig:splay}. The result merged hierarchical tree structure is shallower due to the splay operation.}
\label{fig:splay-merge}
\end{figure}

We augment the merge procedure of $\mathrm{root}(i)$ with a \textbf{splay} operation that reattaches $i$ itself as a direct child of the root $r_i$ by setting $H[i] = r_i$. We illustrate the splay operation in Figure \ref{fig:splay} and \ref{fig:splay-merge} on the same example used in previous section.

This decreases the depth of $i$, while maintaining $\mathrm{root}(i)$ invariant. After a splay, complexity of sequential queries to $\mathrm{root}(i)$ is reduced to $O[1]$. Put in simpler terms, we do not ensure that every leaf of the smaller proto-feature becomes a direct leaf of the root of the bigger feature (as the GADGET linked list implementation), but we maintain that the last accessed point becomes a direct leaf of the root. What we are exploiting here is that pair enumeration yields the same point many times, and hence this reduction implies that the average depth of the hierarchical tree structure is $O[1]$.

However, for a malicious (specially constructed) data set, this optimization can instead lead to significantly worse wall clock times. The hierarchical merging tree $H$ is unbalanced, and can reduce to a linked list. As an illustration, consider the final step where we need to update $H[i]$ to $\mathrm{root}(i)$ for all points in the data set. If the tree is reduced to a linked list, and the iteration starts from the deepest point, the complexity becomes $O[N^2]$. 

To enhance the robustness of the algorithm, we introduce a \textbf{safe guard} operation \footnote{commonly known as path suppression} during splay, by reattaching every parent accessed while back-tracing $H[i]$ as an immediate child of the root. This does not change the scaling of the root tracing and splay operation, but improves the reduction in the depth of the hierarchical tree structure. The complexity of previous malicious example drops from $O[N^2]$ to $O[N]$. In fact, it can be shown that the average time complexity to form a feature of length $L$ in any data set for this algorithm is $O[L\alpha(L)]$ where $\alpha$ is a slowly increasing function, with a value of $4$ for $L=2^{2^{2^{65536}}}$. \citep{Cormen:2009:IAT:1614191} The splay operation thus achieves $O[L]$ complexity without adding additional book-keeping cost in space and time. We find that for cosmological data set, enabling the guard indeed bounds the average depth of the tree closer to $O[1]$.

The pseudo-code for our FOF algorithm along with the splay optimization is listed in Algorithm \ref{alg:fof}. For completeness, we include a proof in the appendix that the pair enumeration algorithm indeed gives the correct solution. 

\begin{algorithm}
\caption{Dual tree Friends-of-Friends with splay.}
\label{alg:fof}
\begin{algorithmic}
\Ensure{$H[i] = H[j]$ iif $i$ and $j$ belongs to the same FOF halo.}
\Procedure{splay}{$H$, $i$}
\Comment{find the subtree root $r$ of a particle $i$, and reattach the particle as an immediate leaf of the root.}
\State{$r \Leftarrow i$}
\While{$H[r] \neq r$}
    \State{$r \Leftarrow H[r]$}
\EndWhile
\Comment{Optionally: add safe-guard by setting $H[i] = r$ along the full path.}
\State{\Return{$H[i] \Leftarrow r$}}
\EndProcedure

\Procedure{merge}{$H$, $i$, $j$}
\Comment{merge the two subtrees containing $i$ and $j$ respectively}
\State{$r \Leftarrow \Call{splay}{H, i}$}
\State{$s \Leftarrow \Call{splay}{H, j}$}
\State{$H[r] \Leftarrow s$}
\EndProcedure

\Procedure{fof}{$H$, $L$, $R$, $P$, $b$}
\State{$H[0:N] \Leftarrow 0 \dots N - 1 $}
\ForAll{$i, j \in \Call{enum}{L, R, P, 0, 0, b}$}
	\State{$\Call{merge}{H, i, j}$}
\EndFor
\For{$i \in 0 \dots N - 1$}
	\State{$\Call{splay}{H, i}$}
\EndFor
\EndProcedure
\end{algorithmic}
\end{algorithm}

\section{Optimization for self-connected KD-tree nodes}

The pair enumeration algorithm produces a large number of pairs in highly over-dense regions such as the cores of halos. We can write down the total number of pairs in a cell of size $b$(linking length) per side with an over-density of $\delta$,
\begin{equation}
N_{op} = [b^3(1 + \delta)]^2 .
\end{equation}
$N_{op}$ can be huge. For example, a typical halo profile predicts an over-density of $\delta \sim 10^6$ at 1 Kpc \citep[e.g.][]{2012MNRAS.422.3081M}. In a high resolution simulation that resolves sub Kpc structures, a typical length of $b = 0.2$ leads to $N_{op} \sim 10^{12}$ operations inside the density peak. 
We will show an optimization that significantly reduces the number of merge operations in very dense regions. For this, we begin with the definition of self-connectedness of nodes. 

\begin{definition}
A KD-Tree node is \textbf{self-connected} when the node $m$ is smaller than the linking length $b$, $\mathrm{MaxDist}(m, m) < b$. A sufficient condition is the maximum distance bound is smaller than the linking length $\mathrm{MaxDistB}(m, m) < b$.
\end{definition}

If a node $m$ is self-connected, then any pair of points $(i, j) \in P[m]$ is separated by at most the linking length $b$. The set of points in $P[m]$, or alternatively node $m$, forms a proto-feature. This proto-feature can be quickly created before the dual tree algorithm by enumerating the points within self-connected nodes, at a cost of $O[b^3(1 + \delta)]$. 

During the dual tree algorithm, we only need to visit and merge a single pair between any two self-connected nodes, due to Theorem \ref{thm:connected}.  This optimization eliminates all but one merge operation of neighboring pairs between any two neighboring self-connected nodes.

We can also reduce the number of distance computations between two fully connected nodes. We apply a simple heuristic to speed up the naive algorithm. We start the enumeration with an approximated closest pair of points, each of which is closest to the center of the other node. Determining the pair only needs $O(N)$ distance computations. We find the heuristics and can quickly terminate the pair enumeration in dense regions, where self-connected nodes tend to be merged. There is a fast algorithm that can determine the exact closest pair between two nodes in $O(N \log N)$ complexity. \citep{4567872} However, the constant time factor is large, making it inefficient for small self-connected nodes. We leave it as future work to incorporate the exact closest pair algorithm on pairs of larger self-connected nodes.

\begin{theorem}
Merging a single neighboring pair from two self-connected nodes is equivalent to visiting and merging all pairs.
\label{thm:connected}
\begin{proof}
By definition, in fully connected nodes $m$ and $n$, there is $a, b$, such that the roots $r_u=r_a$ for any $u \in P[m]$, and $r_v = r_b$ for any $v \in P[n]$.
Therefore, regardless the neighboring pair $(u, v)$ being visited, the only proto-features that are merged are $r_a$ and $r_b$.
\end{proof}
\end{theorem}

We describe the optimized algorithm in Algorithm \ref{alg:connectedenum} and \ref{alg:connectedfof}. The additional array $F[m]$ is non-zero if and only if $m$-th node is self-connected. We use a pre-orderly traversal to mark self-connected nodes. During the traversal, we avoid the formation of proto-feature on the children of a self-connect node with variable $f$.

\begin{algorithm}
\caption{Pair enumeration aware of self-connected nodes.}
\label{alg:connectedenum}
\begin{algorithmic}
\Procedure{connect}{$H$, $P$, $F$, $m$, $b$, $f$}
\If{$f == 0$}
    \If{$\Call{MaxDistB}{m} < b$}
	    \ForAll{$i \in P[m]$}
		    \State{$H[i] \Leftarrow P[m][0]$}
    	\EndFor
	    \State{$F[m] \Leftarrow 1$}
    \EndIf
\Else
\Comment{$m$ must be self-connected because the parent is.}
	\State{$F[m] \Leftarrow 1$}
\EndIf
\State{\Call{connect}{H, P, F, m=L[m], b, f=F[m]}}
\State{\Call{connect}{H, P, F, m=R[m], b, f=F[m]}}
\EndProcedure

\Procedure{enum2}{$H$, $L$, $R$, $P$, $F$, $m$, $n$, $b$}
  	\If{$F[m] = 1$ and $F[n] = 1$}
    	\Comment{connected nodes, terminate after the first neighbour pair}
 		\State{\textbf{yield one} $\Call{enum}{L, R, P, m, n, b}$}
	\EndIf
    \State{\textbf{yield all} $\Call{enum}{L, R, P, m, n, b}$}
\EndProcedure
\end{algorithmic}
\end{algorithm}

\begin{algorithm}
\caption{Friends-of-friends algorithm avoiding self-connected nodes.}
\label{alg:connectedfof}
\begin{algorithmic}
\Ensure{$H[i] = H[j]$ iif $i$ and $j$ belongs to the same FOF halo.}
\Procedure{fof2}{$H$, $F$, $T$, $b$}
\State{$H[0:N] \Leftarrow 0 \dots N - 1 $}
\State{$F[0:\Call{len}{T}] \Leftarrow 0 $}
\State{$\Call{connect}{H, P, F, m=0, b, f=0}$}
\ForAll{$i, j \in \Call{enum2}{H, L, R, P, F, m=0, n=0, b}$}
	\State{$\Call{merge}{H, i, j}$}
\EndFor
\For{$i \in 0 \dots N - 1$}
	\State{$\Call{splay}{H, i}$}
\EndFor
\EndProcedure
\end{algorithmic}
\end{algorithm}

\section{Benchmarks}
In this section we show the performance of our algorithm on two cosmological simulation data sets, each with a total of 16 million points.
\begin{itemize}
\item[Low] A FastPM Simulation in a box of 100 Mpc/h per side with $256^3$ particles. FastPM is a Particle Mesh cosmological simulation code. \citep{2016arXiv160300476F}
\item[High] An NyX Simulation in a box of 10 Mpc/h per side with $256^3$ particles. NyX is an AMR cosmological simulation code. \citep{2013ApJ...765...39A}
\end{itemize}

The tests are performed with a single computing core on a computing node of the Cray XC-30 super-computer Edison at National Energy Research Scientific Computing Center, with a single core on a 2.5 G Ivy-Bridge processor. For each simulation, we perform two sets of runs: one with and one without the safe-guard operation. We use 10 different linking lengths, spanning from 0.01 to 1.0 in units of the mean particle separation. We define the number of book-keeping operations per visit as the total number of operations manipulating the hierarchy or linked list per visit of pairs.

\begin{figure*}
\includegraphics[width=\textwidth]{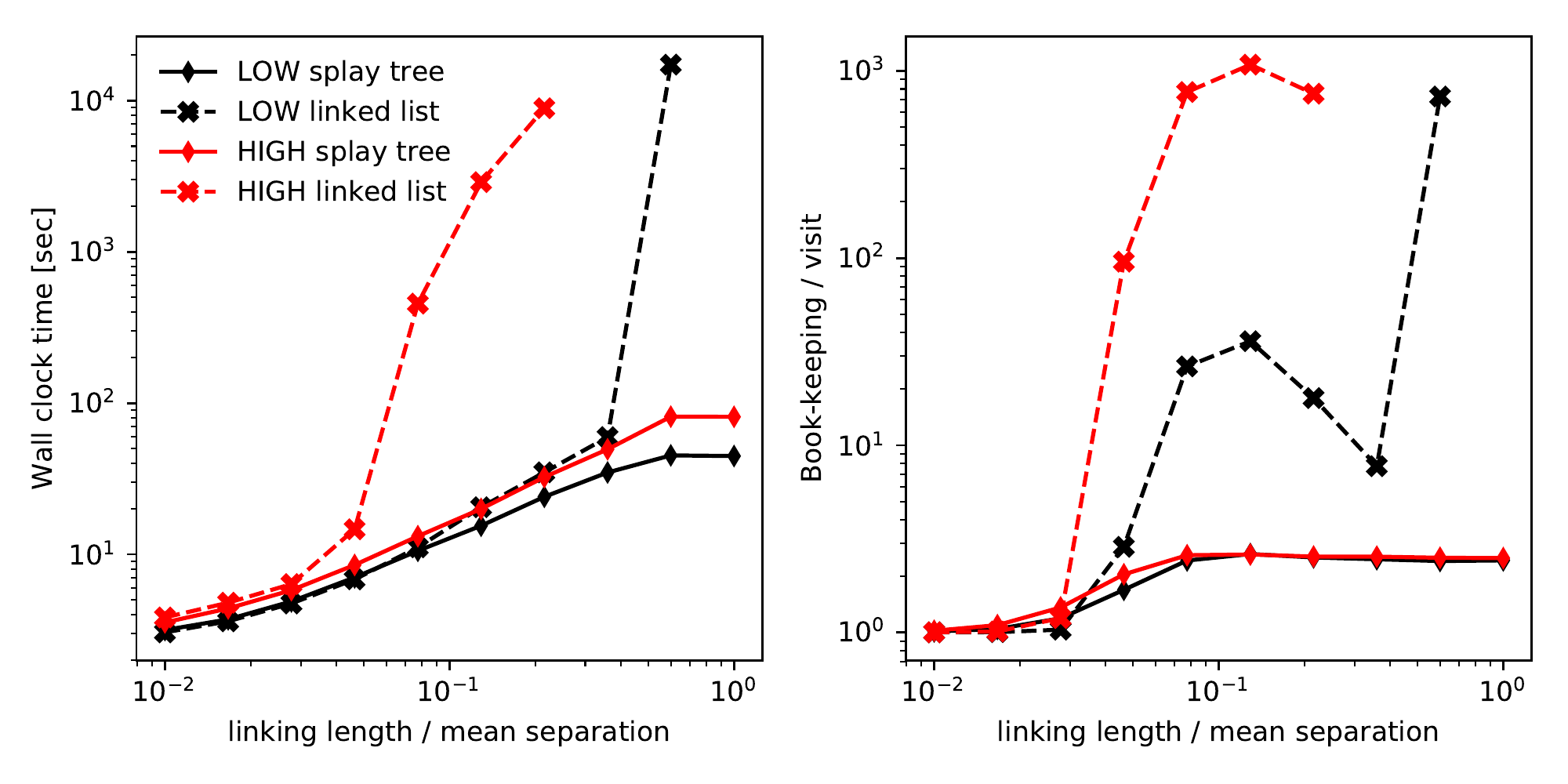}
\caption{Comparing splay tree and linked list implementation. Left: wall clock time. Right: average number of book keeping operations per merging with linked list / splay implementations. Crosses: linked list. (Runs taking more than 3 hours are not presented). Diamonds: splay tree (this work). Red : High resolution dataset, Black: Low resolution dataset. Notice that the splay algorithm is already 30\% faster than linked list at $b = 0.2$ for the low resolution data. For the high resolution data (where features are larger), the splay algorithm is several orders of magnitude faster.}
\label{fig:ll}
\end{figure*}

\begin{figure*}
% nersc m779/biasder/testkdcount/
\includegraphics[width=\textwidth]{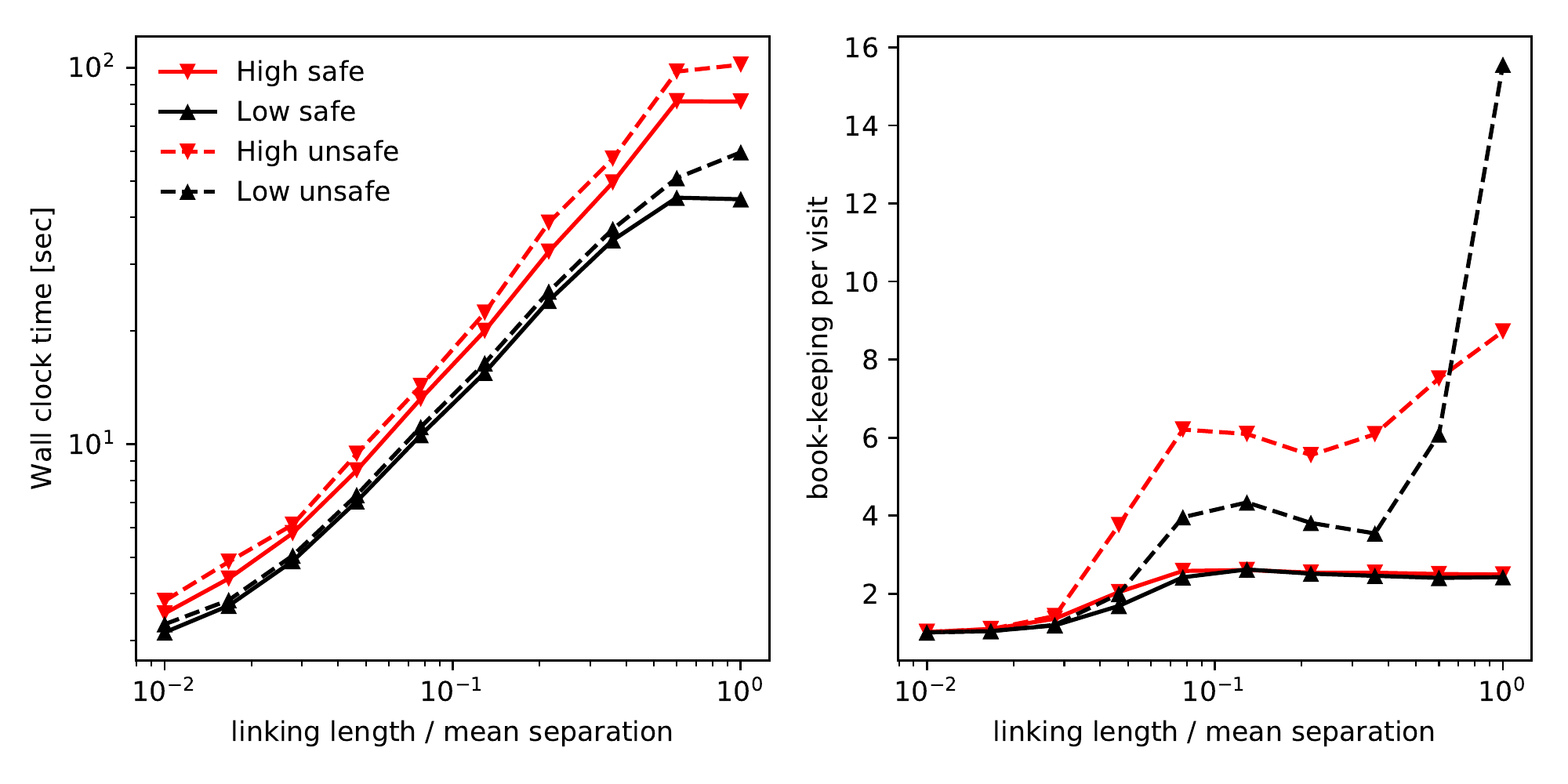}
\caption{Benchmarks comparing the safe guard operation.
Left : wall clock time.
Right : book-keeping operations per visit.
We show the wall clock time and book-keeping cost per visit as a function of linking length. Two data sets are used: the Low resolution FastPM simulation data set (black) and the high resolution NyX simulation data set (red). We show the benchmarks with and without the safe guard operation (dashed / solid) defined in Section 3. While the wall clock time is insensitive to the guard, the number of book keeping operations is indeed bound.
}
\label{fig:results}
\end{figure*}

In Figure \ref{fig:ll}, we investigate the performance due to the choice of representation of the proto-feature data structure. We compare our algorithm with a splay tree against the linked list data structure (which is used in GADGET). For short linking length $b < 0.1$ where most resulted features are small, the algorithms showed little difference, since the time spent in maintaining the hierarchy is negligible. However, as the size of the features increase with $b$, the number of book-keeping operations per visit increases rapidly to the order of hundreds, while the number of book-keeping operations due to the splay tree remains almost a constant up to a few (also seen in the right panel of Figure \ref{fig:results}). As a result, the linked list representation performs considerably worse than the splay tree as the size of features grow. For the low resolution test data set we used, when $b>0.4$, the size of the largest halo increases to more than 5 million points, and the computation of the linked list code could not finish in 20 minutes; while the splay tree representation finished in less than 50 seconds. The linked list algorithm performs much worse on the high resolution data set, being two orders of magnitude slower than the splay algorithm at $b=0.2$. We expect the difference to be more drastic in data sets with even larger halos.

Our algorithm shows robust performance: when the resolution (inverse of mean particle separation) increases by a factor of 10, the wall clock time increases only by a factor of 30 percent. In Figure \ref{fig:results}, we empirically confirm that the safe guard operation bounds the number of book of the hierarchical tree structure per visit. The safe guard operation limits the the average number of book-keeping operations is bound to slightly more than 3. The unsafe operation increases this number of more than 10. The safe guard operation does not affect the wall clock time for the typical data set we tested.

\begin{figure*}
\includegraphics[width=\textwidth]{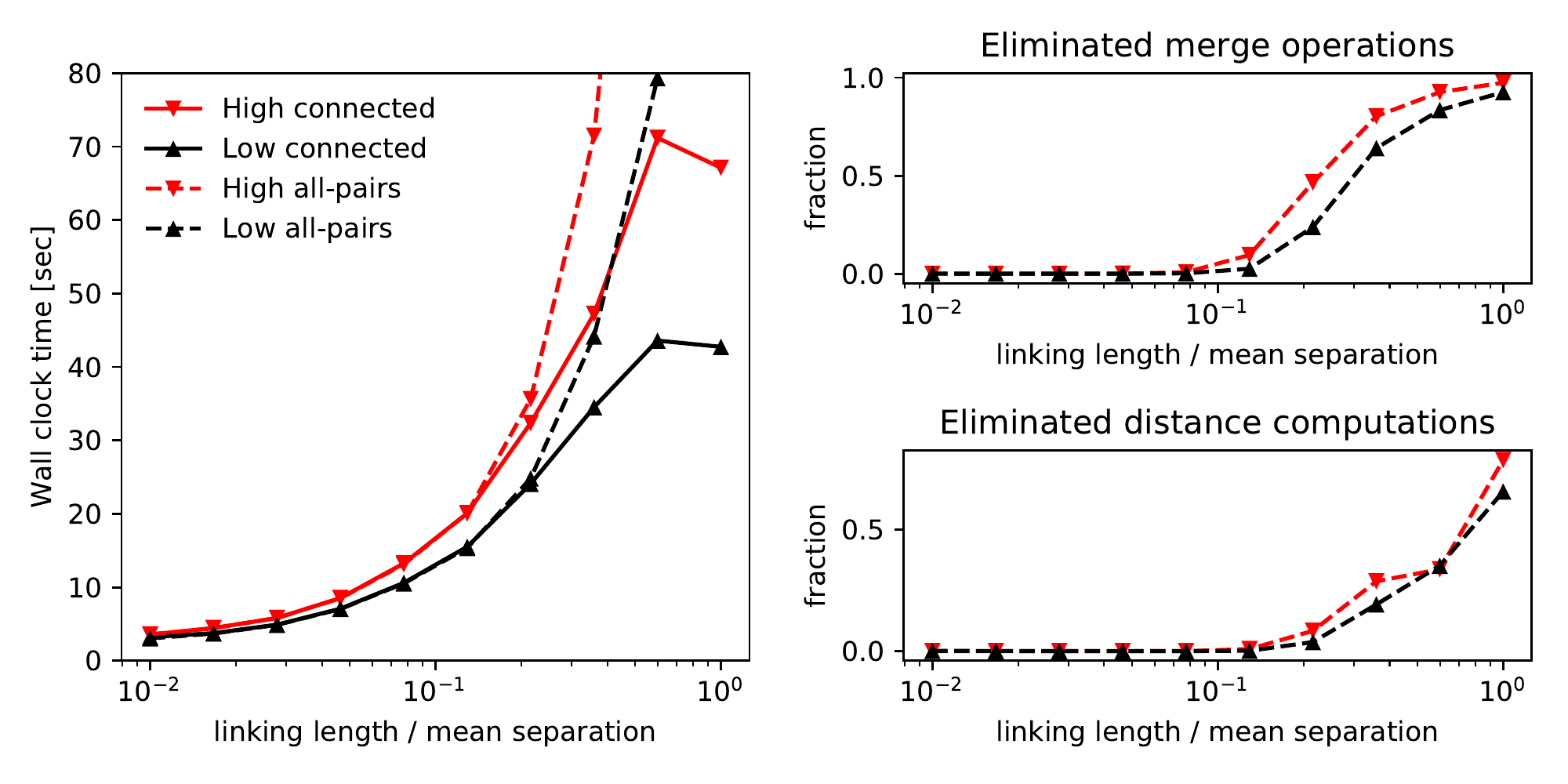}
\caption{Comparing self-connected optimization. 
Left: wall clock time. Upper right: Fraction of merge operations that are eliminated due to the self-connected optimization. Lower right: Fraction of distance computations that are eliminated due to the self-connected optimization.
Solid: with self-connected optimization; Dashed : without self-connected optimization. 
Red : High resolution, Black: Low resolution. The self-connected optimization eliminates about 40\% of the merge operations for the high resolution data at $b=0.2$, and quickly saturates to 1.0. The reduction in number of distance computation is not as drastic at $b=0.2$. abouts 10\% and increases to about 70\% for very large linking lengths. }
\label{fig:connected}
\end{figure*}

In Figure \ref{fig:connected}, we show that as the linking length increases, the self-connected optimization reduces the number of pair enumerations and merge operations by a large fraction, comparing to a without the optimization. At linking length of 0.2, the reduction rate is a factor of $\sim0.5$ for the high resolution data set and about 20\% for the low resolution data set. For extremely large linking length, the reduction rate approaches 1, which could be the reason for the flat wall clock time at very large linking length.

For linking length of 0.2, the algorithm takes less than 20 seconds to process 16 million points ($256^3$) for both high and low resolution simulations. These numbers are comparable to the amount of time spent in reading in the data and preparing the initial KD-Tree (6 seconds). 
Still, there is a huge space for fine tuning, as our KD-Tree implementation is slightly slower than cKDTree in scipy, comparable to halotools \citep{2016arXiv160604106H}, twice slower than TreeCorr \citep{2004MNRAS.352..338J} and 20 times slower than CorrFunc. \citep[an amazingly fast chaining-mesh code] {manodeep_sinha_2016_55161}. We expect a large margin of improvement in speed by adapting a fine-tuned implementation for the base pair enumeration code. 

\section{Conclusion}
We describe a fast algorithm for identifying friends-of-friends halos in cosmological data sets. The algorithm is defined on pair enumeration which visits all edges of the connected graph induced by the linking length and is constructed on a dual KD-tree correlation function code. We present two optimizations that significantly improves the speed and robustness of the FOF algorithm - use of a splay tree and pruning the enumeration of self-connected KD-tree nodes - both of which can be very easily ported to any of the existing pair-enumeration codes. After these two optimizations we find that our algorithm reduces the number of operations for constructing friends of friends halos by almost two orders of magnitude comparing to a naive implementation with linked list for generally used linking lengths of $b=0.2$. 

We began by pointing out that although the friends-and-friends problem is identical to the maximum connected component problem in graph theory, with spatial data such as a cosmological simulation, the elements of adjacent matrix are implied via expensive neighbor queries. Therefore, the dual-tree pair enumeration algorithm that we use is advantageous because it systematically eliminates expensive neighbor queries by tracking two tree nodes simultaneously.

We implement two important optimizations to improve the scaling and robustness of the algorithm for merging proto-features against input data. The first optimization is to append a splay operation to the root query in the hierarchical tree structure of proto-features. The splay operation significantly reduces the average number of traverses, making root-query a $O[1]$ process, without requiring significant additional storage space. The second optimization is to skip merge operations while visiting pairs in two self-connected KD-Tree nodes. Reducing the number of merge operations and pair enumerations significantly speeds up the algorithm in high over-density regions and with large linking lengths. We also proved the correctness of this optimization. 

We note that in our application, the time spent in local friends-of-friends finding becomes sub-dominant comparing to the time in global merging of the catalog. We plan to investigate an optimal distributed algorithm by combining our algorithm with the fast distributed memory parallel algorithm by \citep[e.g.][]{Fu:2010:DDS:1851476.1851527}.

Finally, as an advantage due to insisting on constructing the algorithm with an abstract hierarchical pair enumeration operation, we expect further improvement of speed from our naive implementation by porting the algorithm to a highly optimized correlation function code beyond KD-Tree \citep[e.g.][]{manodeep_sinha_2016_55161}. 

\section*{Acknowledgment}
The authors thank Zarija Lukic for providing the high resolution NyX simulation data set, Ur\v os Seljak for comments on the arrangement of the paper, Martin White for discussions over the heritage of friends-of-friends algorithms. YF thanks Andrew Hearing and Manodeep Sinha for bench marking the performance of the correlation function code implementation used in this work against scipy, halotools, and CorrFunc during the SnowPac 2016 conference in Snowbird, Utah. 
We thank the referee for helping us to identify a bug in the earlier version of the implementation.

\section*{References}

\bibliography{mybibfile}

\begin{thebibliography}{26}
\expandafter\ifx\csname natexlab\endcsname\relax\def\natexlab#1{#1}\fi
\providecommand{\url}[1]{\texttt{#1}}
\providecommand{\href}[2]{#2}
\providecommand{\path}[1]{#1}
\providecommand{\DOIprefix}{doi:}
\providecommand{\ArXivprefix}{arXiv:}
\providecommand{\URLprefix}{URL: }
\providecommand{\Pubmedprefix}{pmid:}
\providecommand{\doi}[1]{\href{http://dx.doi.org/#1}{\path{#1}}}
\providecommand{\Pubmed}[1]{\href{pmid:#1}{\path{#1}}}
\providecommand{\bibinfo}[2]{#2}
\ifx\xfnm\relax \def\xfnm[#1]{\unskip,\space#1}\fi
%Type = Article
\bibitem[{{Almgren} et~al.(2013){Almgren}, {Bell}, {Lijewski}, {Luki{\'c}} and
  {Van Andel}}]{2013ApJ...765...39A}
\bibinfo{author}{{Almgren}, A.S.}, \bibinfo{author}{{Bell}, J.B.},
  \bibinfo{author}{{Lijewski}, M.J.}, \bibinfo{author}{{Luki{\'c}}, Z.},
  \bibinfo{author}{{Van Andel}, E.}, \bibinfo{year}{2013}.
\newblock \bibinfo{title}{{Nyx: A Massively Parallel AMR Code for Computational
  Cosmology}}.
\newblock \bibinfo{journal}{\apj} \bibinfo{volume}{765}, \bibinfo{pages}{39}.
\newblock \DOIprefix\doi{10.1088/0004-637X/765/1/39},
  \href{http://arxiv.org/abs/1301.4498}{\tt arXiv:1301.4498}.
%Type = Article
\bibitem[{{Behroozi} et~al.(2013){Behroozi}, {Wechsler} and
  {Wu}}]{2013ApJ...762..109B}
\bibinfo{author}{{Behroozi}, P.S.}, \bibinfo{author}{{Wechsler}, R.H.},
  \bibinfo{author}{{Wu}, H.Y.}, \bibinfo{year}{2013}.
\newblock \bibinfo{title}{{The ROCKSTAR Phase-space Temporal Halo Finder and
  the Velocity Offsets of Cluster Cores}}.
\newblock \bibinfo{journal}{\apj} \bibinfo{volume}{762}, \bibinfo{pages}{109}.
\newblock \DOIprefix\doi{10.1088/0004-637X/762/2/109},
  \href{http://arxiv.org/abs/1110.4372}{\tt arXiv:1110.4372}.
%Type = Article
\bibitem[{{Bryan} et~al.(2014){Bryan}, {Norman}, {O'Shea}, {Abel}, {Wise},
  {Turk}, {Reynolds}, {Collins}, {Wang}, {Skillman}, {Smith}, {Harkness},
  {Bordner}, {Kim}, {Kuhlen}, {Xu}, {Goldbaum}, {Hummels}, {Kritsuk}, {Tasker},
  {Skory}, {Simpson}, {Hahn}, {Oishi}, {So}, {Zhao}, {Cen}, {Li} and {Enzo
  Collaboration}}]{2014ApJS..211...19B}
\bibinfo{author}{{Bryan}, G.L.}, \bibinfo{author}{{Norman}, M.L.},
  \bibinfo{author}{{O'Shea}, B.W.}, \bibinfo{author}{{Abel}, T.},
  \bibinfo{author}{{Wise}, J.H.}, \bibinfo{author}{{Turk}, M.J.},
  \bibinfo{author}{{Reynolds}, D.R.}, \bibinfo{author}{{Collins}, D.C.},
  \bibinfo{author}{{Wang}, P.}, \bibinfo{author}{{Skillman}, S.W.},
  \bibinfo{author}{{Smith}, B.}, \bibinfo{author}{{Harkness}, R.P.},
  \bibinfo{author}{{Bordner}, J.}, \bibinfo{author}{{Kim}, J.h.},
  \bibinfo{author}{{Kuhlen}, M.}, \bibinfo{author}{{Xu}, H.},
  \bibinfo{author}{{Goldbaum}, N.}, \bibinfo{author}{{Hummels}, C.},
  \bibinfo{author}{{Kritsuk}, A.G.}, \bibinfo{author}{{Tasker}, E.},
  \bibinfo{author}{{Skory}, S.}, \bibinfo{author}{{Simpson}, C.M.},
  \bibinfo{author}{{Hahn}, O.}, \bibinfo{author}{{Oishi}, J.S.},
  \bibinfo{author}{{So}, G.C.}, \bibinfo{author}{{Zhao}, F.},
  \bibinfo{author}{{Cen}, R.}, \bibinfo{author}{{Li}, Y.},
  \bibinfo{author}{{Enzo Collaboration}}, \bibinfo{year}{2014}.
\newblock \bibinfo{title}{{ENZO: An Adaptive Mesh Refinement Code for
  Astrophysics}}.
\newblock \bibinfo{journal}{\apjs} \bibinfo{volume}{211}, \bibinfo{pages}{19}.
\newblock \DOIprefix\doi{10.1088/0067-0049/211/2/19},
  \href{http://arxiv.org/abs/1307.2265}{\tt arXiv:1307.2265}.
%Type = Book
\bibitem[{Cormen et~al.(2009)Cormen, Leiserson, Rivest and
  Stein}]{Cormen:2009:IAT:1614191}
\bibinfo{author}{Cormen, T.H.}, \bibinfo{author}{Leiserson, C.E.},
  \bibinfo{author}{Rivest, R.L.}, \bibinfo{author}{Stein, C.},
  \bibinfo{year}{2009}.
\newblock \bibinfo{title}{Introduction to Algorithms, Third Edition}.
\newblock \bibinfo{edition}{3rd} ed., \bibinfo{publisher}{The MIT Press}.
%Type = Article
\bibitem[{{Davis} et~al.(1985){Davis}, {Efstathiou}, {Frenk} and
  {White}}]{1985ApJ...292..371D}
\bibinfo{author}{{Davis}, M.}, \bibinfo{author}{{Efstathiou}, G.},
  \bibinfo{author}{{Frenk}, C.S.}, \bibinfo{author}{{White}, S.D.M.},
  \bibinfo{year}{1985}.
\newblock \bibinfo{title}{{The evolution of large-scale structure in a universe
  dominated by cold dark matter}}.
\newblock \bibinfo{journal}{\apj} \bibinfo{volume}{292},
  \bibinfo{pages}{371--394}.
\newblock \DOIprefix\doi{10.1086/163168}.
%Type = Article
\bibitem[{{Eisenstein} and {Hut}(1998)}]{1998ApJ...498..137E}
\bibinfo{author}{{Eisenstein}, D.J.}, \bibinfo{author}{{Hut}, P.},
  \bibinfo{year}{1998}.
\newblock \bibinfo{title}{{HOP: A New Group-Finding Algorithm for N-Body
  Simulations}}.
\newblock \bibinfo{journal}{\apj} \bibinfo{volume}{498},
  \bibinfo{pages}{137--142}.
\newblock \DOIprefix\doi{10.1086/305535},
  \href{http://arxiv.org/abs/astro-ph/9712200}{\tt arXiv:astro-ph/9712200}.
%Type = Article
\bibitem[{{Feng} et~al.(2016){Feng}, {Chu} and {Seljak}}]{2016arXiv160300476F}
\bibinfo{author}{{Feng}, Y.}, \bibinfo{author}{{Chu}, M.Y.},
  \bibinfo{author}{{Seljak}, U.}, \bibinfo{year}{2016}.
\newblock \bibinfo{title}{{FastPM: a new scheme for fast simulations of dark
  matter and halos}}.
\newblock \bibinfo{journal}{ArXiv e-prints}
  \href{http://arxiv.org/abs/1603.00476}{\tt arXiv:1603.00476}.
%Type = Inproceedings
\bibitem[{Fu et~al.(2010)Fu, Ren, L\'{o}pez, Fink and
  Gibson}]{Fu:2010:DDS:1851476.1851527}
\bibinfo{author}{Fu, B.}, \bibinfo{author}{Ren, K.},
  \bibinfo{author}{L\'{o}pez, J.}, \bibinfo{author}{Fink, E.},
  \bibinfo{author}{Gibson, G.}, \bibinfo{year}{2010}.
\newblock \bibinfo{title}{Discfinder: A data-intensive scalable cluster finder
  for astrophysics}, in: \bibinfo{booktitle}{Proceedings of the 19th ACM
  International Symposium on High Performance Distributed Computing},
  \bibinfo{publisher}{ACM}, \bibinfo{address}{New York, NY, USA}. pp.
  \bibinfo{pages}{348--351}.
\newblock \URLprefix \url{http://doi.acm.org/10.1145/1851476.1851527},
  \DOIprefix\doi{10.1145/1851476.1851527}.
%Type = Article
\bibitem[{{Geller} and {Huchra}(1983)}]{1983ApJS...52...61G}
\bibinfo{author}{{Geller}, M.J.}, \bibinfo{author}{{Huchra}, J.P.},
  \bibinfo{year}{1983}.
\newblock \bibinfo{title}{{Groups of galaxies. III - The CfA survey}}.
\newblock \bibinfo{journal}{\apjs} \bibinfo{volume}{52},
  \bibinfo{pages}{61--87}.
\newblock \DOIprefix\doi{10.1086/190859}.
%Type = Article
\bibitem[{{Hearin} et~al.(2016){Hearin}, {Campbell}, {Tollerud}, {Behroozi},
  {Diemer}, {Goldbaum}, {Jennings}, {Leauthaud}, {Mao}, {More}, {Parejko},
  {Sinha}, {Sipocz} and {Zentner}}]{2016arXiv160604106H}
\bibinfo{author}{{Hearin}, A.}, \bibinfo{author}{{Campbell}, D.},
  \bibinfo{author}{{Tollerud}, E.}, \bibinfo{author}{{Behroozi}, P.},
  \bibinfo{author}{{Diemer}, B.}, \bibinfo{author}{{Goldbaum}, N.J.},
  \bibinfo{author}{{Jennings}, E.}, \bibinfo{author}{{Leauthaud}, A.},
  \bibinfo{author}{{Mao}, Y.Y.}, \bibinfo{author}{{More}, S.},
  \bibinfo{author}{{Parejko}, J.}, \bibinfo{author}{{Sinha}, M.},
  \bibinfo{author}{{Sipocz}, B.}, \bibinfo{author}{{Zentner}, A.},
  \bibinfo{year}{2016}.
\newblock \bibinfo{title}{{High-Precision Forward Modeling of Large-Scale
  Structure: An open-source approach with Halotools}}.
\newblock \bibinfo{journal}{ArXiv e-prints}
  \href{http://arxiv.org/abs/1606.04106}{\tt arXiv:1606.04106}.
%Type = Article
\bibitem[{{Hopkins} et~al.(2014){Hopkins}, {Kere{\v s}}, {O{\~n}orbe},
  {Faucher-Gigu{\`e}re}, {Quataert}, {Murray} and
  {Bullock}}]{2014MNRAS.445..581H}
\bibinfo{author}{{Hopkins}, P.F.}, \bibinfo{author}{{Kere{\v s}}, D.},
  \bibinfo{author}{{O{\~n}orbe}, J.}, \bibinfo{author}{{Faucher-Gigu{\`e}re},
  C.A.}, \bibinfo{author}{{Quataert}, E.}, \bibinfo{author}{{Murray}, N.},
  \bibinfo{author}{{Bullock}, J.S.}, \bibinfo{year}{2014}.
\newblock \bibinfo{title}{{Galaxies on FIRE (Feedback In Realistic
  Environments): stellar feedback explains cosmologically inefficient star
  formation}}.
\newblock \bibinfo{journal}{\mnras} \bibinfo{volume}{445},
  \bibinfo{pages}{581--603}.
\newblock \DOIprefix\doi{10.1093/mnras/stu1738},
  \href{http://arxiv.org/abs/1311.2073}{\tt arXiv:1311.2073}.
%Type = Article
\bibitem[{{Jarvis} et~al.(2004){Jarvis}, {Bernstein} and
  {Jain}}]{2004MNRAS.352..338J}
\bibinfo{author}{{Jarvis}, M.}, \bibinfo{author}{{Bernstein}, G.},
  \bibinfo{author}{{Jain}, B.}, \bibinfo{year}{2004}.
\newblock \bibinfo{title}{{The skewness of the aperture mass statistic}}.
\newblock \bibinfo{journal}{\mnras} \bibinfo{volume}{352},
  \bibinfo{pages}{338--352}.
\newblock \DOIprefix\doi{10.1111/j.1365-2966.2004.07926.x},
  \href{http://arxiv.org/abs/astro-ph/0307393}{\tt arXiv:astro-ph/0307393}.
%Type = Article
\bibitem[{{Koda} et~al.(2016){Koda}, {Blake}, {Beutler}, {Kazin} and
  {Marin}}]{2016MNRAS.459.2118K}
\bibinfo{author}{{Koda}, J.}, \bibinfo{author}{{Blake}, C.},
  \bibinfo{author}{{Beutler}, F.}, \bibinfo{author}{{Kazin}, E.},
  \bibinfo{author}{{Marin}, F.}, \bibinfo{year}{2016}.
\newblock \bibinfo{title}{{Fast and accurate mock catalogue generation for
  low-mass galaxies}}.
\newblock \bibinfo{journal}{\mnras} \bibinfo{volume}{459},
  \bibinfo{pages}{2118--2129}.
\newblock \DOIprefix\doi{10.1093/mnras/stw763},
  \href{http://arxiv.org/abs/1507.05329}{\tt arXiv:1507.05329}.
%Type = Inproceedings
\bibitem[{Kwon et~al.(2010)Kwon, Nunley, Gardner, Balazinska, Howe and
  Loebman}]{Kwon:2010:SCA:1876037.1876051}
\bibinfo{author}{Kwon, Y.}, \bibinfo{author}{Nunley, D.},
  \bibinfo{author}{Gardner, J.P.}, \bibinfo{author}{Balazinska, M.},
  \bibinfo{author}{Howe, B.}, \bibinfo{author}{Loebman, S.},
  \bibinfo{year}{2010}.
\newblock \bibinfo{title}{Scalable clustering algorithm for n-body simulations
  in a shared-nothing cluster}, in: \bibinfo{booktitle}{Proceedings of the 22Nd
  International Conference on Scientific and Statistical Database Management},
  \bibinfo{publisher}{Springer-Verlag}, \bibinfo{address}{Berlin, Heidelberg}.
  pp. \bibinfo{pages}{132--150}.
\newblock \URLprefix \url{http://dl.acm.org/citation.cfm?id=1876037.1876051}.
%Type = Article
\bibitem[{{Liu} et~al.(2008){Liu}, {Hsieh}, {Ho}, {Lin} and
  {Yan}}]{2008ApJ...681.1046L}
\bibinfo{author}{{Liu}, H.B.}, \bibinfo{author}{{Hsieh}, B.C.},
  \bibinfo{author}{{Ho}, P.T.P.}, \bibinfo{author}{{Lin}, L.},
  \bibinfo{author}{{Yan}, R.}, \bibinfo{year}{2008}.
\newblock \bibinfo{title}{{A New Galaxy Group Finding Algorithm: Probability
  Friends-of-Friends}}.
\newblock \bibinfo{journal}{\apj} \bibinfo{volume}{681},
  \bibinfo{pages}{1046--1057}.
\newblock \DOIprefix\doi{10.1086/588183}.
%Type = Article
\bibitem[{{Martizzi} et~al.(2012){Martizzi}, {Teyssier}, {Moore} and
  {Wentz}}]{2012MNRAS.422.3081M}
\bibinfo{author}{{Martizzi}, D.}, \bibinfo{author}{{Teyssier}, R.},
  \bibinfo{author}{{Moore}, B.}, \bibinfo{author}{{Wentz}, T.},
  \bibinfo{year}{2012}.
\newblock \bibinfo{title}{{The effects of baryon physics, black holes and
  active galactic nucleus feedback on the mass distribution in clusters of
  galaxies}}.
\newblock \bibinfo{journal}{\mnras} \bibinfo{volume}{422},
  \bibinfo{pages}{3081--3091}.
\newblock \DOIprefix\doi{10.1111/j.1365-2966.2012.20879.x},
  \href{http://arxiv.org/abs/1112.2752}{\tt arXiv:1112.2752}.
%Type = Inproceedings
\bibitem[{{Moore} et~al.(2001){Moore}, {Connolly}, {Genovese}, {Gray}, {Grone},
  {Kanidoris}, {Nichol}, {Schneider}, {Szalay}, {Szapudi} and
  {Wasserman}}]{2001misk.conf...71M}
\bibinfo{author}{{Moore}, A.W.}, \bibinfo{author}{{Connolly}, A.J.},
  \bibinfo{author}{{Genovese}, C.}, \bibinfo{author}{{Gray}, A.},
  \bibinfo{author}{{Grone}, L.}, \bibinfo{author}{{Kanidoris}, II, N.},
  \bibinfo{author}{{Nichol}, R.C.}, \bibinfo{author}{{Schneider}, J.},
  \bibinfo{author}{{Szalay}, A.S.}, \bibinfo{author}{{Szapudi}, I.},
  \bibinfo{author}{{Wasserman}, L.}, \bibinfo{year}{2001}.
\newblock \bibinfo{title}{{Fast Algorithms and Efficient Statistics: N-Point
  Correlation Functions}}, in: \bibinfo{editor}{{Banday}, A.J.},
  \bibinfo{editor}{{Zaroubi}, S.}, \bibinfo{editor}{{Bartelmann}, M.} (Eds.),
  \bibinfo{booktitle}{Mining the Sky}, p.~\bibinfo{pages}{71}.
\newblock \DOIprefix\doi{10.1007/10849171_5},
  \href{http://arxiv.org/abs/astro-ph/0012333}{\tt arXiv:astro-ph/0012333}.
%Type = Article
\bibitem[{{Murphy} et~al.(2012){Murphy}, {Geach} and
  {Bower}}]{2012MNRAS.420.1861M}
\bibinfo{author}{{Murphy}, D.N.A.}, \bibinfo{author}{{Geach}, J.E.},
  \bibinfo{author}{{Bower}, R.G.}, \bibinfo{year}{2012}.
\newblock \bibinfo{title}{{ORCA: The Overdense Red-sequence Cluster
  Algorithm}}.
\newblock \bibinfo{journal}{\mnras} \bibinfo{volume}{420},
  \bibinfo{pages}{1861--1881}.
\newblock \DOIprefix\doi{10.1111/j.1365-2966.2011.19782.x},
  \href{http://arxiv.org/abs/1109.3182}{\tt arXiv:1109.3182}.
%Type = Misc
\bibitem[{{Nbody-Shop}(a)}]{AFOF}
\bibinfo{author}{{Nbody-Shop}}, a.
\newblock \bibinfo{title}{{Approximated Friends-of-Friends}}.
\newblock \URLprefix
  \url{http://www-hpcc.astro.washington.edu/tools/afof.html}.
%Type = Misc
\bibitem[{{Nbody-Shop}(b)}]{KDFOF}
\bibinfo{author}{{Nbody-Shop}}, b.
\newblock \bibinfo{title}{{KD-Tree Friends-of-Friends}}.
\newblock \URLprefix \url{http://www-hpcc.astro.washington.edu/tools/fof.html}.
%Type = Inproceedings
\bibitem[{Shamos and Hoey(1975)}]{4567872}
\bibinfo{author}{Shamos, M.I.}, \bibinfo{author}{Hoey, D.},
  \bibinfo{year}{1975}.
\newblock \bibinfo{title}{Closest-point problems}, in: \bibinfo{booktitle}{16th
  Annual Symposium on Foundations of Computer Science (sfcs 1975)}, pp.
  \bibinfo{pages}{151--162}.
\newblock \DOIprefix\doi{10.1109/SFCS.1975.8}.
%Type = Article
\bibitem[{Shun and Blelloch(2013)}]{Shun:2013:LLG:2517327.2442530}
\bibinfo{author}{Shun, J.}, \bibinfo{author}{Blelloch, G.E.},
  \bibinfo{year}{2013}.
\newblock \bibinfo{title}{Ligra: A lightweight graph processing framework for
  shared memory}.
\newblock \bibinfo{journal}{SIGPLAN Not.} \bibinfo{volume}{48},
  \bibinfo{pages}{135--146}.
\newblock \URLprefix \url{http://doi.acm.org/10.1145/2517327.2442530},
  \DOIprefix\doi{10.1145/2517327.2442530}.
%Type = Misc
\bibitem[{Sinha(2016)}]{manodeep_sinha_2016_55161}
\bibinfo{author}{Sinha, M.}, \bibinfo{year}{2016}.
\newblock \bibinfo{title}{Corrfunc: Corrfunc-1.1.0}.
\newblock \URLprefix \url{http://dx.doi.org/10.5281/zenodo.55161},
  \DOIprefix\doi{10.5281/zenodo.55161}.
%Type = Article
\bibitem[{Sleator and Tarjan(1985)}]{Sleator:1985:SBS:3828.3835}
\bibinfo{author}{Sleator, D.D.}, \bibinfo{author}{Tarjan, R.E.},
  \bibinfo{year}{1985}.
\newblock \bibinfo{title}{Self-adjusting binary search trees}.
\newblock \bibinfo{journal}{J. ACM} \bibinfo{volume}{32},
  \bibinfo{pages}{652--686}.
\newblock \URLprefix \url{http://doi.acm.org/10.1145/3828.3835},
  \DOIprefix\doi{10.1145/3828.3835}.
%Type = Article
\bibitem[{{Springel}(2005)}]{2005MNRAS.364.1105S}
\bibinfo{author}{{Springel}, V.}, \bibinfo{year}{2005}.
\newblock \bibinfo{title}{{The cosmological simulation code GADGET-2}}.
\newblock \bibinfo{journal}{\mnras} \bibinfo{volume}{364},
  \bibinfo{pages}{1105--1134}.
\newblock \DOIprefix\doi{10.1111/j.1365-2966.2005.09655.x},
  \href{http://arxiv.org/abs/astro-ph/0505010}{\tt arXiv:astro-ph/0505010}.
%Type = Article
\bibitem[{{White} et~al.(2010){White}, {Cohn} and {Smit}}]{2010MNRAS.408.1818W}
\bibinfo{author}{{White}, M.}, \bibinfo{author}{{Cohn}, J.D.},
  \bibinfo{author}{{Smit}, R.}, \bibinfo{year}{2010}.
\newblock \bibinfo{title}{{Cluster galaxy dynamics and the effects of
  large-scale environment}}.
\newblock \bibinfo{journal}{\mnras} \bibinfo{volume}{408},
  \bibinfo{pages}{1818--1834}.
\newblock \DOIprefix\doi{10.1111/j.1365-2966.2010.17248.x},
  \href{http://arxiv.org/abs/1005.3022}{\tt arXiv:1005.3022}.

\end{thebibliography}
\appendix
\section{Proof of the algorithm}
Theorem \ref{thm:correct} guarantees the algorithm finds a solution to the maximum connected component problem. 
\begin{theorem}
The features in $H$ after the algorithm are fully connected and maximum: $H[i] = H[j]$ if and only if $i$, $j$ is connected by a path ( belongs to the same feature)
\label{thm:correct}
\begin{proof}
We first prove that if $root(i) = root(j)$, then $i$ and $j$ are connected. We can show this by induction. At the beginning of the algorithm, this is clearly true, since there is no $i \neq j$ such that $root(i) = root(j)$. Without losing generality we consider the case where a visit to $i, j$ sets $H[r_j] = r_i$, and the claim is true before the visit. Denote the points in the two old subtrees $A(r_i)$ and $A(r_j)$, then by the assumption, $A(r_i)$ is connected and $A(r_j)$ is connected. Because $i\in A(r_i)$, $j\in A(r_j)$ and $i, j$ are connected, $A(r_i) \cup A(r_j)$ is also connected. Since $A(r_j)$ are the only points that has changed root, the algorithm maintains the invariant at each iteration. Note that the last step of the algorithm ensures $H[i] = root(i)$. 

Next we prove that if $i$ and $j$ are connected, then $H[i] = H[j]$. We show this by contradiction. Assume there exists point $i$ and $k$, where $H[k] \neq H[i]$, and $Dist(i, k) < b$. Because any $(i, j)$ such that $Dist(i, j) < b$ has been visited, the pair $i, k$ must have been visited during the pair enumeration. A merge operation ensures $j = root(k) = root(i)$. Any later merge operations will only change the root of $j$. Thus $root(k) = root(j)$ from this point on till the end of enumeration, resulting $H[k] = H[i]$. The contradiction shows that there is no such pair $i, k$. Therefore every feature we have identified is maximized. 

Because every feature is fully connected and maximized, the algorithm has identified all friends-of-friends features from the data set.
\end{proof}

\end{theorem}

\end{document}